# Induced smectic ordering and blue phase formation in mixtures of cyanobiphenyls and cholesterol esters

Natalia A. Kasian[a,b], Longin N. Lisetski[a,*], Serhii E. Ivanchenko[c], Vitalii O. Chornous[d], Halyna V. Bogatyryova[e], Igor A. Gvozdovskyy[e,**]

[a]*Department of Nanostructured Materials, Institute for Scintillation Materials of STC "Institute for Single Crystals" of the National Academy of Sciences of Ukraine, Kharkiv, Ukraine;* [b]*Faculty of Physics, Warsaw University of Technology, Warsaw, Poland;* [c]*Department of Physical Chemistry and Technology of Nanostructured Ceramics and Nanocomposites, Frantsevich Institute for Problems of Materials Science of the National Academy of Sciences of Ukraine, Kyiv, Ukraine;* [d]*Department of Medical and Pharmaceutical Chemistry, Bukovinian State Medical University, Chernivtsi, Ukraine;* [d]*Department of Optical Quantum Electronics, Institute of Physics of the National Academy of Sciences of Ukraine, Kyiv, Ukraine;*

Institute for Scintillation Materials of STC "Institute for Single Crystals" of the National Academy of Sciences of Ukraine, 60 Nauky ave., 61072, Kharkiv, Ukraine, telephone number: +380 057 3410321, *E-mail: lcsciencefox@gmail.com

Institute of Physics of the National Academy of Sciences of Ukraine, 46 Nauky ave., Kyiv, 03028, Ukraine, telephone number: +380 44 5250862, **E-mail: igvozd@gmail.com

It has been found that in mixtures of nematic E7 and smectogenic cholesteryl oleyl carbonate (COC) the $S_A$-$N^*$ transition temperature is substantially (by ~20 K) increased, as compared with pure COC, at E7 concentrations around ~40%. Within the same concentration range, the isotropic transition is preceded by formation of blue phase, with its maximum width of ~3.5 K clearly correlated to the increased thermal stability of the $S_A$ phase. With other cholesterol esters or cyanobiphenyls (cholesteryl nonanoate, 5CB), this effect was either much weaker or not observed. No enhancement of smectic phase was noted when E7 was replaced by $N_{tb}$-forming nematic mixture based on CB7CB with two cyanobiphenyl moieties. Selective Bragg reflection of light (BRL) spectra were measured in all three temperature regions, including the unwinding of the cholesteric helix on cooling towards S$_A$ phase and characteristic selective BRL changes in the blue phase. In the latter case, the measured λ$_{max}$ values were dependent both on the helical twisting power in the cholesteric phase and on the lattice size and orientation in the blue phase. Also considered were the effects of ferroelectric BaTiO$_3$ nanoparticles accumulated at disclination lines upon the blue phase thermal stability.

## 1.  Introduction

Among specific phenomena that can be observed in multi-component liquid crystal systems, one of the most intriguing is formation of induced smectic phases (ISP) in nematic mixtures comprising components of different chemical nature. In this case, certain features of intermolecular interactions evoke substantial enhancement of the tendency to translational ordering, which is already present in latent form in the component molecules.

The most frequently noted mechanism giving rise to ISP is the formation of charge-transfer complexes, *e.g.*, between alkylcyanobiphenyls and azomethines (Schiff bases), as described in a classical work of Park et al. [1] However, other possible mechanisms were also proposed, together with a review of other pioneering works in this field. [2] No evidence of intermolecular complex formation was found for 5CB doped with 4-pentylbenzoic acid system manifesting clearly expressed ISP properties; [3] on the other hand, formation of complexes between alkylcyanobiphenyl and dialkylazoxybenzene molecules was clearly noted even in isotropic solvents. [4] A comprehensive review and analysis for various cases of ISP formation was presented in. [5] Since then, there has been not much attention to this topic. An attempt to develop a generalised theoretical model describing ISP formation caused by complexation via specific interactions between two dissimilar molecules was made in. [6] Combining the Maier-Saupe theory for orientational ordering and McMillan theory for smectic layered ordering with the Flory-Huggins theory for isotropic mixing, the authors succeeded in deriving various types of phase diagrams observed in different systems. One can also mention detailed studies of the 4-n-octyl-4´-cyanobiphenyl (8CB) and 4,4´-di-n-heptyl-azobenzene (7AB7), forming a great variety of induced smectic phases upon mixing. [7]

As a more recent example, ISP was noted in mixtures of two nematogenic substances possessing low-temperature twist-bend phases (a cyanobiphenyl dimer and benzoyloxy-benzylidene dimer) [8]. In this case, the mixtures showed no new absorption bands in the wide spectral range, implicating the absence of any significant charge transfer or even dipole - induced dipole interaction. The smectic arrangement was presumably promoted by geometrical factors, with intercalated structures of the constituent molecules providing the necessary driving force for ISP.

Another interesting case of ISP formation due to geometrical factors was noted in [9]. In the mixtures of 5CB and cholesteryl oleyl carbonate (COC), the $S_A$-$N^*$ transition temperature of COC was increased by about 3 K upon addition of 20 wt. % of 5CB, and returned to its initial value only at ~ 40 wt. % of 5CB. The explanation was given in terms of 5CB monomers or dimers (depending on 5CB concentration) aligning themselves between the stretched long hydrocarbon chains of adjacent COC molecules.

However, no less intriguing and interesting is another feature that can be observed in highly chiral media, namely, the formation of so-called "blue phases" (BPs), which are thermodynamically stable within a narrow temperature range (*e.g.* about 1 °C, as experimentally found by Meiboom et al [10] for cholesteryl nonanoate (CN)) .The existence of BPs was first theoretically explained by Brazovskii et al. [11,12] Later Meiboom et al [13] and Wright et al [14] have described, basing on the Oseen-Frank elasticity equations, an unstable planar texture of cholesterics ($N^*$) near the isotropic phase (Iso) characterised by a network of disclination lines of body-centered cubic (*i.e.* BPI) or simple cubic (*i.e.* BPII) lattice (as schematically shown in Figure 1 (a), (b) and Figure 1 (c), (d), respectively).

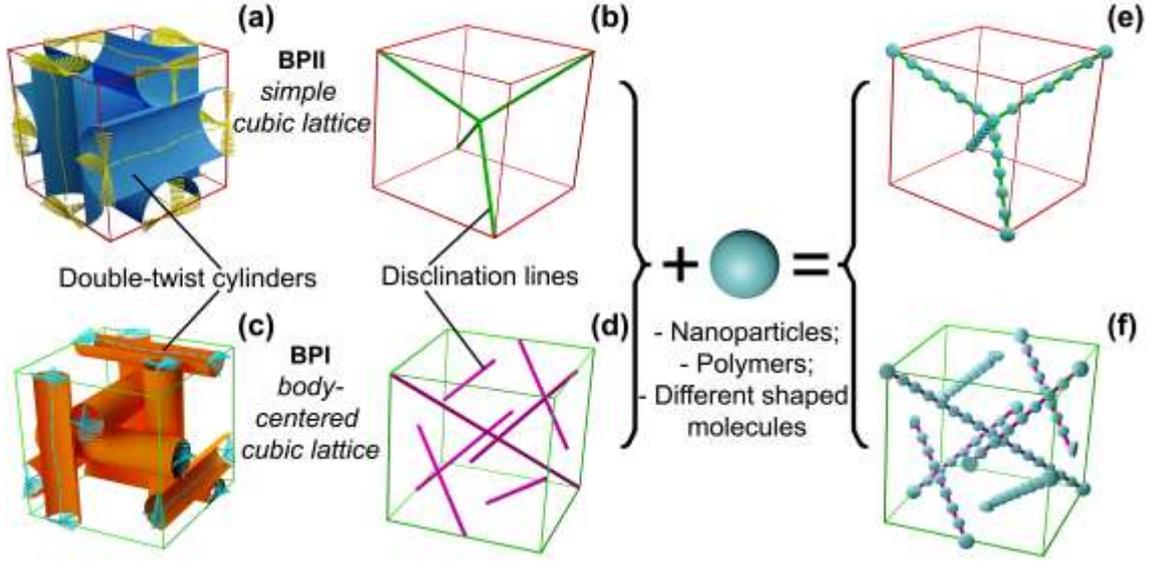

Figure 1. Schematic presentation of BPs characterised by: (a) simple cubic lattice for BPII and (c) body-centered cubic lattice for BPI formed by double-twist cylinders. BPs disclination lines are shown for (b) BPII and (d) BPI. Stabilisation of the BPs by nanoparticles is shown in (e) for BPII and (f) for BPI.

Blue phases as highly chiral liquid crystal media are characterised by the presence of double-twist cylinders (Figure 1 (a), (c)) [14] and self-assembled into 3D photonic crystals showing selective BRL with relatively narrow spectrum. Due to this fact they have high potential of application in the manufacturing technology of displays, various kinds of sensors and photonic devices. [15-23] Such BP media are characterised by typical platelet textures [24] having domains with differently oriented lattice planes (*i.e.* selective BRL colours) observed in polarising optical microscope (POM). The lattice planes are characterised by Miller indices [$h$, $k$, $l$] ($h + k + l$ is an even number), and the reflected light, according to Bragg's law, can be written as follows:

$$\lambda_{(h,k,l)} = \frac{2n \cdot a}{\sqrt{h^2 + l^2 + k^2}} \qquad (1),$$

where $a$ is the lattice constant of BPs and $n$ is refractive index.

Dierking et al have shown in [25,26] that experimentally observed platelet texture of BPs can be rather accurately recreated by using Voronoi diagrams. [27] The uniform alignment of the lattice planes by using different aligning layers (*e.g.* numerically studied by Fukuda et al [28] and experimentally obtained by Glesson et al

[29], Kikuchi et al [30] and Otón et al, [31]) and patterned surfaces, [32-34] leads to clearly observed diffraction lines under the reflective POM (*i.e.* so-called Kossel diagrams). [35,36] This is important for the use of BP medium for different applications.

While high chirality of liquid-crystalline medium is main condition to form BPs, the availability of network of the lattices is even more important condition. This is supported by formation of BPs even when this network is filled by non-chiral (*e.g.* nematic E7) liquid-crystalline medium, as shown by Xiang and Lavrentovich. [37] In addition, with series of non-chiral bent-core dimeric molecules recently discovered by Imrie et al, [38] chiral liquid crystalline mediums can be formed having smectic structure, which in the case of chiral bent-core molecules with cholesteryl group, can additionally possess blue phases. [39,40]

The significant drawback of BPs, *i.e.* their narrow temperature range was largely overcome by adding to liquid-crystalline medium of different types of nanoparticles (NPs), [41,42] polymer mesogens to form polymer networks [43-45] or various molecules having different shapes (*e.g.* bimesogenic, [46] T-shaped, [47] hydrogen-bonded [47-49] and non-chiral bend-core molecules [38]). In addition, it should be noted that the increase of stability of the BPs temperature range could be achieved using surface-functionalised graphene nanosheets – such effect was observed by Lavrič at al [50] even for minute concentrations of dispersed graphene.

In this manuscript the ISP formation in mixture based on nematic E7 doped with cholesteryl oleyl carbonate (COC) within wide concentration range from 0 to 100 wt. % was studied. It has been shown that the blue phase appeared in the same COC concentrations where the ISP was the most pronounced. Further temperature stabilisation of BP was achieved by adding of ferroelectric nanoparticles $BaTiO_3$ as previous described in detail by Wang et al. [51] In addition, alongside E7-COC, we studied mixtures of COC with twist-bend nematics CB7CB/CB6OCB/5CB in ratio (39:19:42) under similar experimental conditions. Also, in parallel series of experiments we used cholesteryl nonanoate (CN) as another example of smectic-forming cholesterol esters. As compared to COC, its hydrocarbon chain is saturated and much shorter.

## 2. Materials and methods

*2.1 Materials*

To study the induced smectic phase the nematic liquid crystals mixture E7, which contains cyanobiphenyl and cyanoterphenyl components, was obtained from Licrystal, Merck (Darmstadt, Germany). It was chosen because its temperature of the nematic-isotropic transition is higher (*i.e.* $T_{Iso}$ is 58 ºC [52]) than for recently described mixtures on the base of the nematic 5CB. [9] As another type of the nematic liquid crystalline medium to form the ISP, we used a three-component mixture consisting of two achiral liquid crystal dimers, *i.e.* 1'-,7"-bis-4-(4-cyanobiphen-4'-yl)heptane (CB7CB) and 1-(4-cyanobiphenyl-4'-yloxy)-6-(4-cyanobiphenyl-4'-yl)hexane (CB6OCB), and the monomer nematic 4-pentyl-4'-cyanobiphenyl (5CB) (Figure 2 (a), (b), (c)) in weight ratio (39:19:42), which was recently used to prepare $N^*_{tb}$-forming system possessing the oblique helicoidal structure. [53,54] The twist-bend nematic CB7CB was synthesised in Bukovinian State Medical University (Chernivtsi, Ukraine) with its characteristics described in detail. [55] The N phase of synthesised CB7CB is observed for temperature range from 103 to 116 °C. The twist-bend nematic CB6OCB was obtained from Synthon Chemicals GmbH & Co (Wolfen, Germany), and possesses N phase within temperature range from 109 to 157 ºC. [55-57] The nematic 5CB was synthesised in STC "Institute of Single Crystals" (Kharkiv, Ukraine) and purified before using.

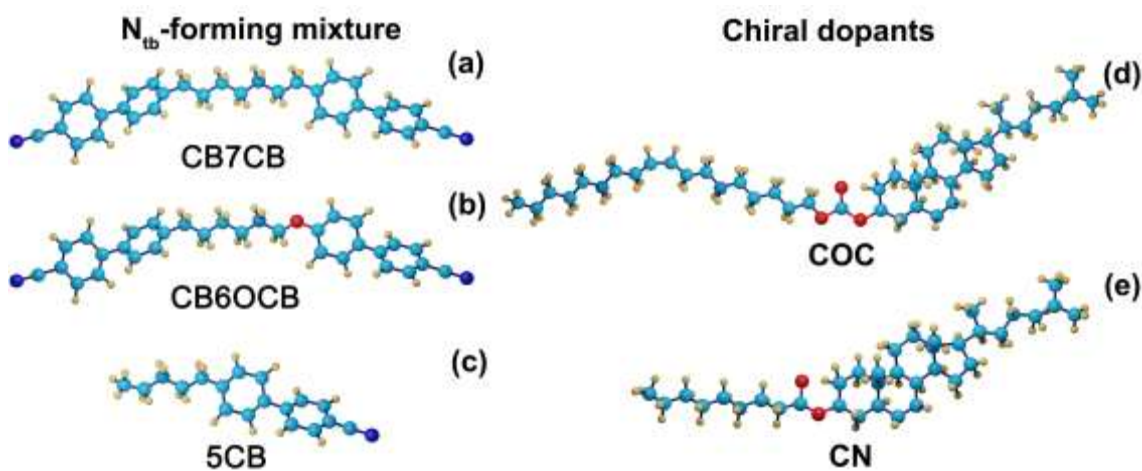

Figure 2. Chemical structures of compounds used. The twist-bend nematic-forming mixture consists of: (a) - 39 wt. % of 1'-,7"-bis-4-(4-cyanobiphen-4'-yl)heptane (CB7CB), (b) - 19 wt. % of 1-(4-cyanobiphenyl-4'-yloxy)-6-(4-cyanobiphenyl-4'-yl)hexane (CB6OCB) and (c) - 42 wt. % of 4-pentyl-4'-cyanobiphenyl (5CB). The cholesterol esters as chiral compounds: (d) - cholesteryl oleyl carbonate (COC) and (e) - cholesteryl nonanoate (CN).

The cholesteryl oleyl carbonate (COC) (Figure 2 (d)) as chiral dopant (ChD) was obtained from Aldrich (USA). COC possesses various liquid-crystalline phases, including smectic A phase ($S_A$), the temperatures of phases transitions and their ranges were presented in detail by Lebovka at al. [9]

The cholesteryl nonanoate (CN) (Figure 2 (e)) was synthesised in A.V. Bogatsky Physico-Chemical Institute of the NAS of Ukraine (Odesa, Ukraine). By taking into account the fact that for CN the $S_A$ phase is also observed [58,59], we also used this substance for a comparative study.

The concentration of ChDs in the nematic host E7 for both the COC and the CN was changed in range from 0 to 100 wt. %. The concentration of ChDs in ISP was changed with the step about 5 wt. %.

To stabilise BPs of COC-containing mixtures, we added ferroelectric $BaTiO_3$ NPs synthesized in Frantsevich Institute for Problems of Materials Science of the NAS of Ukraine (Kyiv, Ukraine). The size of NPs with an average diameter of approximately 24 nm was achieved. The NPs were covered by oleic acid and further dissolved in heptane to create uniformed suspension. This suspension was mixed with LC mixture, containing the chiral dopant COC and nematic E7 in different weight ratio. The detailed description of preparation of ferroelectric NPs is given in the section Methods.

To obtain planar alignment of various studied mixtures (*i.e.* $N^*$ and $N^*_{tb}$-forming mixtures) polyimide PI2555 (HD MicroSystems, USA), [60] known for its strong azimuthal anchoring energy, [61] was used.

*2.2 Methods*

The glass substrates (microscope slides, made in Germany) were covered with n-menthyl-2-pyrrolidone solution of PI2555 in ratio 10:1 and spin-coated (6800 rpm over 10 s) to obtain a thin alignment film. The substrates were dried at 80 °C for 15 min, and subsequent thermopolymerisation of polyimide molecules at 180 °C was carried out during 30 min. The PI2555 film was rubbed $N_{rubb}$ = 15 times to ensure strong azimuthal anchoring energy about $10^{-5}$ J/m$^2$. [61]

LC cells were assembled of two rubbed substrates oriented in such a way that each of them had opposite rubbing direction with respect to another. The thickness of LC cell, set by spherical spacer of 20 µm diameter, was 21 ± 0.5 µm. The thickness was measured by interference method by recording of transmission spectrum of the empty LC cell using an Ocean Optics USB4000 spectrometer (Ocean Insight, USA, California).

The chiral mixtures were prepared within wide concentration ratio of ChDs in host liquid-crystalline medium (*i.e.* N, $N_{tb}$-forming mixture) from 0 to 100 wt. % with the minimum step about 5 wt. %. To create the uniform chiral mixtures, a Fisher Vortex Genie 2 mixer (Fisher Scientific, USA) was used with periodically heating the mixture to Iso (vortex 4-5 over 30 min). The LC cell was filled by using the capillary method at the temperature just above the isotropic transition of chiral mixtures to avoid the additional aligning of LC owing to the flow.

To stabilise the temperature range of BPs, as was described in detail elsewhere, [51] we used the ferroelectric nanosized barium titanate ($BaTiO_3$) powder mainly consisting of NPs with ~ 24 nm in diameter. The NPs of $BaTiO_3$ were synthesised using the non-isothermal decomposition method of high-purity barium titanyl oxalate, *i.e.* $Ba(TiO)(C_2O_4)_2$, produced by Degussa Electronic Corporation (Netherlands). The non-isothermal decomposition of barium titanyl oxalate was carried out in a rotary furnace with 14 temperature zones, developed at Frantsevich Institute for Problems of Materials Science of the NAS of Ukraine (Kyiv, Ukraine), at temperatures within 650 - 700 °C.

To prepare the dispersion, BaTiO3 nanopowder (Nanotechcenter LLC, Ukraine) with a mean particle diameter of 24 nm was used. The powder was submerged in heptane (Heltermann Carless, Germany) with the addition of oleic acid (Heltermann Carless, Germany) at a mass ratio of 1:1.8:13.3, respectively. The dispersion was ground in a 50 ml glass jar with zirconia oxide balls for 48 hours on a rotary mill at a speed of 100 rpm.

The liquid-crystalline suspensions were prepared by doping of the LC mixture (*e.g.* COC-E7 mixture containing 72.5 wt. % of COC and 27.5 wt. % of E7 or CN-E7 mixture containing 62.3 wt. % of CN and 31.7 wt. % of E7) by $BaTiO_3$ NPs within range from 0.25 to 4 wt. %. The subsequent 30 min sonication of the suspension at temperature of BP-I was carried out by means of the UZDN-2T ultrasonic disperser (Ukrrospribor, Sumy, Ukraine), at the frequency of 22 kHz and the output power of 75 W.

To measure the phase transition temperatures of studied mixtures, we used a thermostable heater based on a temperature regulator MikRa 603 (LLD 'MikRa', Kyiv, Ukraine) additionally equipped with a platinum resistance thermometer Pt1000 (PJSC 'TERA', Chernihiv, Ukraine). The temperature measurement accuracy was ± 0.1 °C. During the BPs phase transition studies the speed of temperature change was about 0.01 °C/min (0.1 °C per 10 min) both on heating and cooling.

### 3. Results and discussions

*3.1. Experimental study of induced smectic phase mixtures based on cholesteryl oleyl carbonate and nematic E7*

This section will be a certain continuation of studies of the induced smectic phase in systems based on non-smectogenic nematic 5CB and cholesteryl oleyl carbonate (COC) recently described in detail by Lebovka at al. [9] Two important features were noted for these mixtures, namely first of all, the increase of temperature of the cholesteric to smectic A ($S_A$) transition induced in COC by addition of up to 20 wt. % of 5CB and secondly, the linear dependence of reciprocal helix pitch $1/P$ on concentration of nematic 5CB, as distinct from the most cases of nematic-cholesteric mixtures. However, the use of nematic 5CB to study all the concentration range from 0 to 100 wt. % in this ISP mixture is somewhat limited due to the low temperature of the Iso-N transition ($T_{Iso}$ = 35.5 °C).

To increase the concentration range of the non-smectogenic nematic in ISP mixture in this work we propose to use the nematic characterised by higher temperature of Iso-N transition, for instance, the nematic E7 ($T_{Iso}$ = 58 °C [52]) which includes 51 wt. % of 5CB.

The phase transitions temperatures of the ISP mixtures, based on nematic E7 and certain concentration $C_{COC}$ of smectogenic chiral dopant COC, which lies in the range of 0 - 100 wt. %, are shown for both on heating (Figure 3a) and on cooling (Figure 3b). The Figure 3 also shows that during heating and cooling process the certain hysteresis is observed.

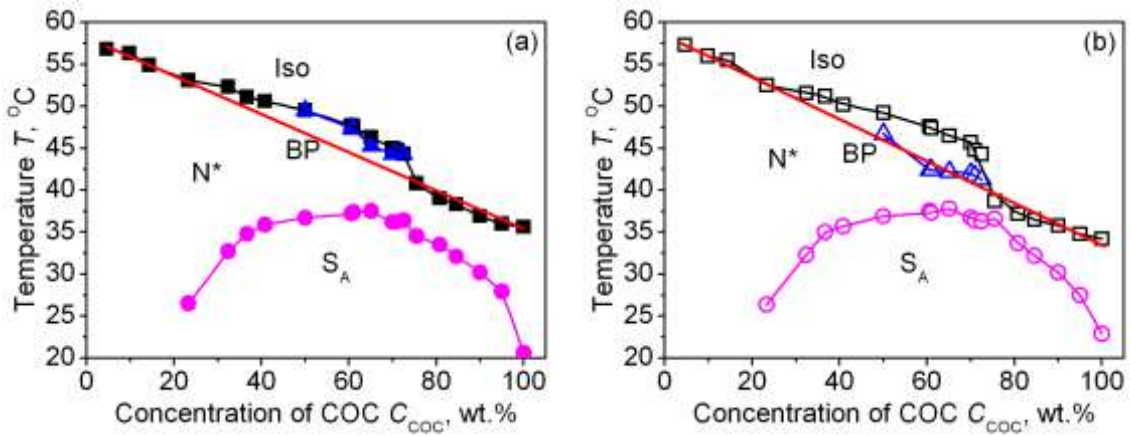

Figure 3. Dependence of temperatures of the phase transitions of ISP mixtures, based on COC and E7, on concentration $C_{COC}$ of the chiral dopant COC during: (a) heating (solid symbols) and (b) cooling (opened symbols) processes. During heating process the sequential phase transitions of the E7-COC mixtures are the next: $S_A$ – $N^*$ (solid magenta circles), $N^*$ - Iso (solid black squares), $N^*$ - BPs (solid blue triangles), and BPs –Iso (solid black squares). During cooling process the sequential phase transitions of ISP mixtures are the next: Iso – $N^*$ (opened black squares), Iso – BPs (opened black squares), BPs – $N^*$ (opened blue triangles), $N^*$ - $S_A$ (opened magenta circles). The linear approximation for the $N^*$ - Iso transition during both heating and cooling is shown by solid red line.

Figure 4 shows sequential change of textures of phase transitions of ISP mixture, *e.g.*, having COC concentration $C_{COC}$ = 72.5 wt. %, during cooling process. As can be seen from Figure 3b this ISP mixture is characterised by all phases inherent in it, including BPs, $N^*$ and $S_A$.[62]

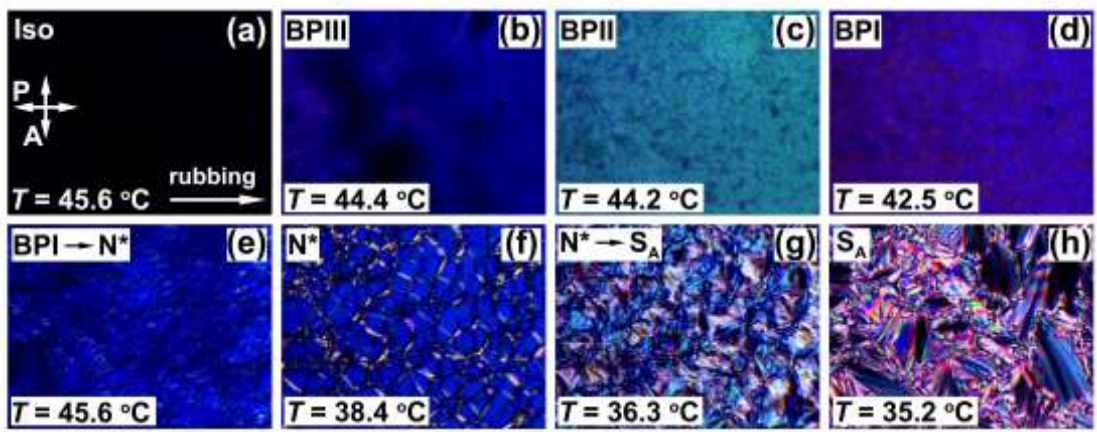

Figure 4. Texture of ISP mixture, containing 72.5 wt. % of COC and 27.5 wt. % of nematic E7 on cooling: (a) Iso at 45.6 °C; (b) Iso – BPIII transition at 44.4 °C; (c) BPII characterised by simple cubic lattice at 44.2 °C; (d) BPI characterised by body-centered cubic lattice at 42.5 °C; (e) BPI - $N^*$ transition at 45.6 °C; (f) $N^*$ phase at 38.4 °C, characterised helicoidal structure, with Grandjean-Cano '*planar*' texture and oily streaks defects; (g) $N^*$ - $S_A$ transition at 36.3 °C and (h) $S_A$ phase at 35.2 °C. Thickness of LC cell was 26.1 µm. LC cell was placed between crossed polariser (P) and analyser (A). The rubbing direction of PI2555 layers was coincided with direction of polariser (P).

The certain deviation from linear dependence of temperature Iso-$N^*$ transition on concentration of COC is observed (Figure 3) as was described in detail elsewhere [9] for ISP mixture, based on COC and 5CB. For the mixtures COC and E7 at concentrations of COC within range from 50 to 75 wt. %, the deviation is noted during both the heating and cooling processes (Figure 3). It can also be seen that at these concentrations the region of blue phases appears additionally in ISP mixtures. It is interesting that t in this range of concentrations the temperatures of the $S_A$ – $N^*$ transitions (magenta circles) are the highest and show only minor variation with temperature (within about 35 – 37 °C). Thus, it appears that there is a certain correlation and probably a relationship between the induction of translational order and formation of the blue phase. This will be discussed in detail in our further narration.

The maximum of wavelength $\lambda_{max}$ of selective BRL in visible range of spectrum as function of temperature *T* for the ISP, consisting of 72.5 wt. % of COC in E7, is shown in Figure 5. The red shift of $\lambda_{max}$ of selective BRL occurs during cooling of $N^*$. For this ISP mixture $N^*$ possesses sufficient shift of wavelength (*i.e.* about 120 nm) when the temperature changes within small range from 36 to 39 °C. This temperature range refers to the human body temperature, including both the normal and high temperature, and can be proposed for biomedical applications using the same ideology as, *e.g.*, in Ref. [63]

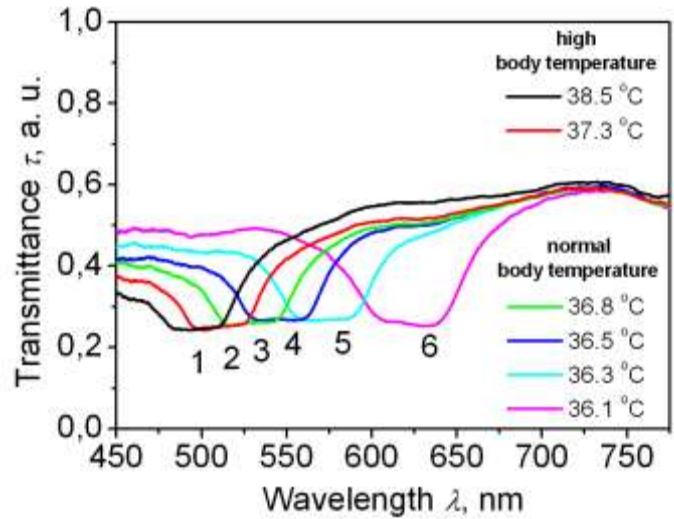

Figure 5. Transmission spectra of the N* containing 72.5 wt. % of COC and 27.5 wt. % of E7 at different temperatures: 1) – 38.5 °C (black spectrum); 2) – 37.3 °C (red spectrum); 3) – 36.8 °C (green spectrum); 4) – 36.5 °C (blue spectrum); 5) – 36.2 °C (cyan spectrum) and 6) – 36.1 °C (magenta spectrum). Thickness of LC cell was 26.1 µm.

Figure 6 shows the dependencies of reciprocal helix pitch $1/P$ on temperature $T$ in the case when ISP mixture containing a certain concentration of COC is in the N* phase. The decrease of temperature $T$ leads to the red shift of maximum wavelength of the selective BRL in the visible range, *i.e.* increasing of the pitch length $P$ of helix in the N* phase.

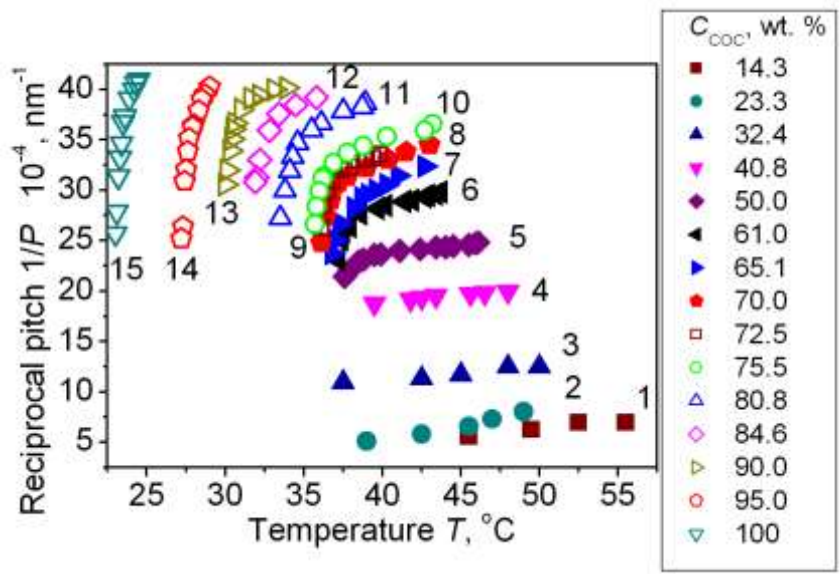

Figure 6. Dependence of reciprocal helix pitch 1/*P* on temperature *T* ISP mixture, containing certain concentration of COC, namely: 1) 14.3 wt. % (solid brown squares), 2) – 23.3 wt. % (solid dark cyan circles), 3) – 32.4 wt. % (solid royal triangles up), 4) – 40.8 wt. % (solid magenta triangles down), 5) – 50.0 wt. % (solid purple diamonds), 6) – 61 wt. % (solid black triangles left), 7) – 65.1 wt. % (solid blue triangles right), 8) – 70 wt. % (solid red pentagons), 9) – 72.5 wt. % (opened brown squares), 10) – 75.5 wt. % (opened green circles), 11) – 80.8 wt. % (opened blue triangles up), 12) – 84.6 wt. % (opened magenta diamonds), 13) - 90.0 wt. % (opened dark yellow triangles right), 14) – 95 wt. % (opened red pentagons), 15 – 100 wt. % (opened dark triangles down).

*3.2 Blue phases of induced smectic phase mixtures based on cholesteryl oleyl carbonate and nematic E7*

In this section we will consider the ISP mixtures possessing BPs. As was mentioned above, the BP is observed within concentration range of COC about 50 – 75 wt. %, where the deviation from linear dependence of $T_{\text{Iso}}(C)$ is observed (Figure 3).

The textures of sequential BPs transitions during cooling of ISP mixture, consisting 72.5 wt. % of COC and 27.5 wt. % of nematic E7, are shown in Figure 4 (b) - (e).

Typically, in contrast to the process of heating (Figure 7(a)) during cooling the temperature ranges of BPs of ISP mixtures were wider (Figure 7(b)).

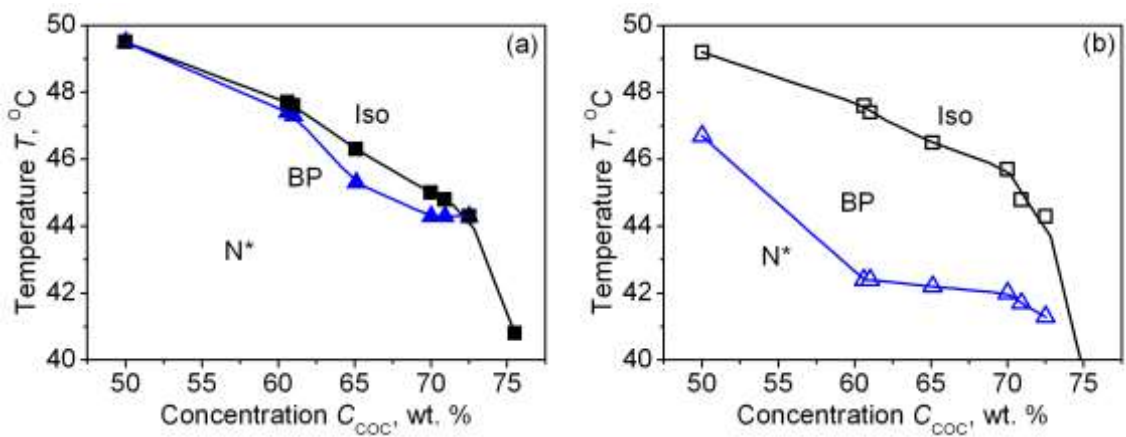

Figure 7. Temperatures of phase transitions of BPs of ISP-forming mixtures E7-COC within COC concentration range from 50 to 75 wt. % on heating (a) and cooling (b).

The transmittance spectrum of one of these BP-forming mixtures (72.5% COC and 27.5% E7) at different temperatures is shown in Figure 8. During cooling from Iso (black spectrum 1) the appearance of the BPIII (red spectrum 2), so-called fog phase, is observed. Further cooling process leads to sequential appearance of BPII (green spectrum 3) and supercooled BPI (blue spectrum 4), characterised by simple cubic (Figure 2 (b)) and body-centered cubic (Figure 2 (c)) lattices, [62] respectively.

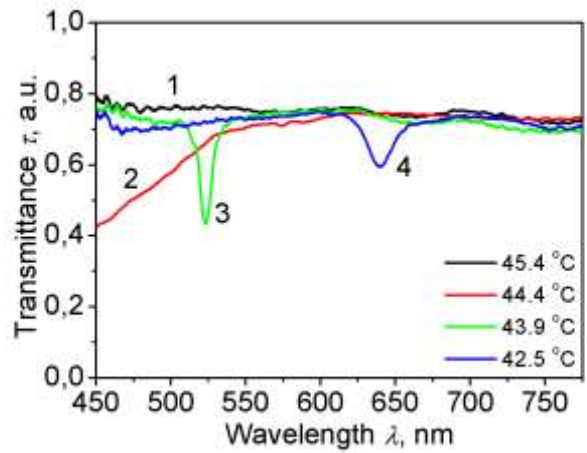

Figure 8. Transmission spectra of the BPs of IPS mixture containing 72.5 wt. % of COC and 27.5 wt. % of E7 at different temperatures during cooling: 1) – Iso at 45.4 °C (black spectrum); 2) – BPIII (*i.e.* fog phase) at 44.4 °C (red spectrum); 3) – BPII at 43.9 °C (green spectrum); 4) – BPI at 42.5 °C (blue spectrum). Thickness of LC cell was 26.1 µm.

The use of ISP mixtures that are additionally characterised by BPs allows us to experimentally observe both continuous changes of maximum wavelength BRL of the $N^*$ at low temperatures (Figure 5) and the discontinuous changes of maximum wavelength BRL of the BPs at high temperatures of mixture (Figure 8).

Because the temperature range of BPs is narrow (Figure 7 (b)), then small temperature changes lead to the sufficient discontinuous shift (*i.e.* about 100 nm) of maximum of the selective BRL wavelength $\lambda_{max}$, as can be seen in Figure 8. The main reason of discontinuous shift is the presence of the lattice for BPs. The selective BRL is described by Equation (1), which relates the selective reflection wavelength to characteristics of the lattice plane with indices of Miller (h,l,k). The relationship between the length of cholesteric helix $P$ (*i.e.* $P \sim (\beta \times C)^{-1}$, where $\beta$ and $C$ is the helical twisting power of ChD and its concentration, respectively [62]) and BPs lattice size $a$ for the first time was experimentally considered by Otón et al., [64] and our results seem to be in a qualitative agreement with their data.

*3.3. Experimental study of induced smectic phase mixtures based on cholesteryl nonanoate and nematic E7*

Our next step was to check whether the rather unusual behavior of correlated appearance of ISP and BPs in the E7-COC system would be observed for other chiral dopants forming $S_A$.

In this section we will consider the mixtures of E7 and cholesteryl nonanoate (CN). It is known that CN possesses sequential phases transitions, namely under heating Cr 75 °C $S_A$ 78.6°C $N^*$ 91.2°C Iso. [59] It should be noted that cholesterol esters from C9 to C14 and some of their mixtures were studied by Duarte et al, [65] paying attention to density measurements during first and second order $S_A$ – $N^*$ phase transitions.

Figure 9 shows, both on heating and cooling, the phase transition temperatures for various mixtures based on nematic E7 and CN possessing the $S_A$ phase. [58,59] Typically, certain small hysteresis of the temperatures during the process of heating and cooling is observed.

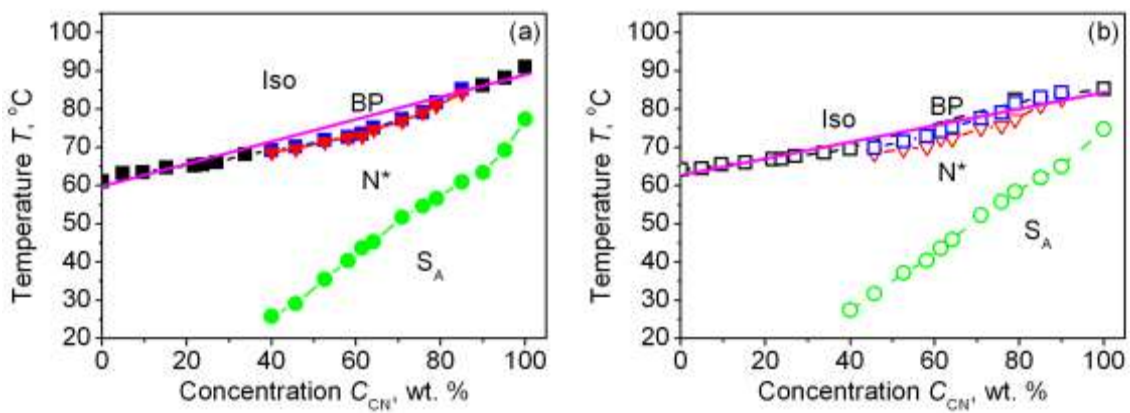

Figure 9. Dependence of temperatures of the phase transitions of E7-CN systems on concentration $C_{CN}$ of the chiral dopant CN during: (a) heating (solid symbols) and (b) cooling (opened symbols) processes. During heating process the sequential phase transitions of the E7-CN mixtures are: $S_A$ – $N^*$ (solid green circles), $N^*$ - Iso (solid black

squares), $N^*$ - BPs (solid red triangles), and BPs – Iso (solid blue squares). During cooling process the sequential phase transitions of ISP mixtures are: Iso – $N^*$ (opened black squares), Iso – BPs (opened blue squares), BPs – $N^*$ (opened red triangles), $N^*$ - $S_A$ (opened green circles). The linear approximation for the $N^*$ - Iso transition during both the process of heating and the cooling is shown by solid magenta line.

As in the case of E7-COC mixtures, the sequential phase transitions including $S_A$, $N^*$, BP and Iso are also observed. The temperatures of $S_A$ - $N^*$, $N^*$ - BP and BPs – Iso transitions are monotonically increasing with the increase of concentration $C_{CN}$ of CN. The dependence of temperature of phase transition on concentration of CN for the Iso (black symbols) has certain deviation from linear function (solid magenta line). This deviation is observed for the ISPs that are characterised by the appearance of BPs. In contrary to ISP mixtures based on COC (Figure 3) the appearance of BPs for CN-containing mixtures within concentration range from 40 to 80 wt. % leads to the decrease of temperature under BP - Iso transition (Figure 9). But the main difference between E7-COC and E7-CN systems is that the latter are, in fact, not ISP systems – only a slight positive deviation from linearity of the $S_A$ - $N^*$ transition temperature is observed. This is accompanied by the fact that for the E7-CN system the observed BPs are noticeably narrower.

The E7-COC mixtures are characterised by their rather pronounced ISP mesomorphism, which for E7-CN is extremely weak. The most probable reason is that ISP with COC is due to steric interactions between molecules of cyanobiphehyl and cholesteryl oleyl carbonate with is very long hydrocarbon chain – this allows a certain equilibrium between cyanobiphenyl dimers and cyanobiphenyl monomers arranged between the tails of adjacent COC molecules (as discussed in detail in [9]). Too short alkyl chains of cholesteryl nonanoate do not allow such arrangement, and ISP formation is not observed.

Also, for mixtures, the obtained result is in good agreement with the density measurements of first and second order phase transitions of cholesterol esters and their mixtures that are characterised by both $S_A$ and BPs. [65]

Finally, one can make an assumption that the observed correlation between ISP and BP formation can be explained by the short-range translational ordering above the $S_A$ - $N^*$ transition, which could stimulate the lattice formation close to the isotropic transition.

*3.4. ISP mixtures possessing stabilised BPs by means of BaTiO$_3$ NPs*

It is well known that the adding of NPs to BPs leads to extension of the temperature interval of their existence. The main reason of stabilisation of BPs is that NPs can fill space in disclinations of lattice of the BPs, as schematically shown in Figure 2 (e), (f).

In this section we will consider the temperatures of phase transitions of ISP mixtures with blue phases stabilised by small concentration of ferroelectric barium titanate (BaTiO$_3$) NPs. The main purpose of these studies is to find out the effects of NPs used to stabilise BPs upon changes in the temperatures of phase transitions, especially BP-Iso transitions for which certain deviations from linear dependence are observed (Figure 3).

It should be noted that NPs of ferroelectric BaTiO$_3$ was previously used for stabilisation of BPs of the mixture containing two chiral compounds, namely R-811 and Iso-(6OBA)$_2$, and the nematic SLC-X. [51]

Figure 10 (a) shows the dependence of phase transition temperatures on concentration of BaTiO$_3$ NPs for the mixtures that are characterised by the wider temperature range of BPs, *e.g.* 72.5 wt.% COC and 27.5 wt.% E7, as can be seen from Figure 3 (b) and Figure 7 (b). The adding of the NPs leads to extension of the temperature range of the BPs.

Thus, for ISP mixture containing BaTiO$_3$ NPs within certain concentration range from 0.5 to 1 wt. % the maximal broadening of the temperature range of existence of the BPs was observed (Figure 10 (b)). For this concentration range a small increase in temperature Iso (solid black squares, curve 1, Figure 10 (a)) that is typical for stabilisation of BPs; in particular, this applies to BaTiO$_3$ NPs described in detail elsewhere. [51]

The Figure 10 (a) shows that the increase of NPs concentration to 4 wt. % of BaTiO$_3$ leads to the decrease of temperature Iso - BPs transition to the value as it is observed for the BPs without adding of NPs. However, the significant broadening of temperature range of BPs due to decrease of temperature transition BPs - N$^*$ (solid blue circles, curve 2) is observed.

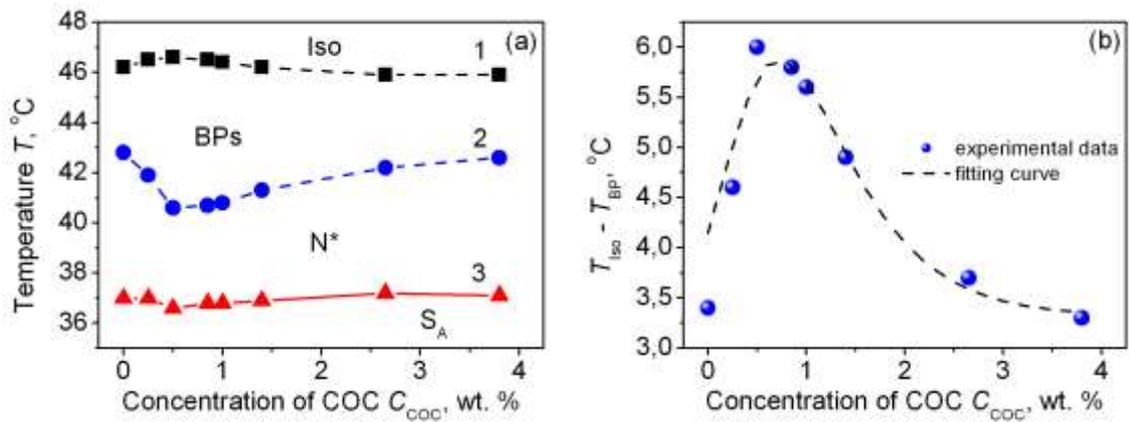

Figure 10. Dependence of the (a) temperature of phase transitions and (b) temperature difference between Iso and BPs during cooling for the ISP mixture consisting of 72.5 wt. % of COC and 27.5 wt. % of nematic E7 on concentration of $BaTiO_3$ NPs.

According to the data shown in Figure 10, we can assume that maximal influence on the deviation from linear dependence under phase transition to Iso should be observed when the presence of the 0.5 wt. % of $BaTiO_3$ NPs in the ISP mixtures having BPs. However, this deviation will be non significant, because the adding of $BaTiO_3$ NPs leads to a small increase of the temperature of the BPs - Iso transition.

Figure 11 shows the dependence of the temperature for Iso - BPs (black squares symbols) and BPs - $N^*$ (blue circles symbols) transitions of ISP mixtures containing certain concentration ratio between COC and E7 and additionally doped by 0.5 wt. % of $BaTiO_3$ NPs, on the COC concentration.

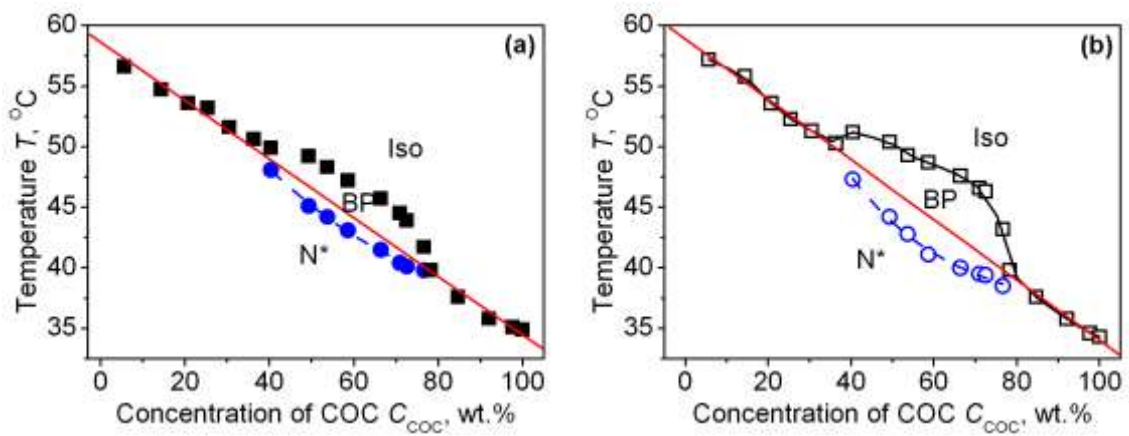

Figure 11. Dependence of the Iso - BP (black squares symbols) and BP - $N^*$ (blue circles symbols) phase transition temperatures in the studied ISP mixtures on concentration of COC during (a) heating (b) and cooling. The E7-COC ISP mixtures

were additionally doped by 0.5 wt. % BaTiO$_3$ NPs. The linear dependence $T(C)$ is shown by solid red line.

By comparing Figure 11 and Figure 3 it is clearly seen that the presence of BaTiO$_3$ NPs in BPs leads to increased deviation from linear dependence of Iso transition (*i.e.*, on other words, the pronounced convexity). In addition, BaTiO$_3$ NPs stabilise BPs, so we can see the minimum hysteresis during heating (Figure 11 (a)) and cooling (Figure 11 (b)), in comparison with the ISP mixtures without NPs (Figure 3).

The adding of BaTiO$_3$ NPs to mixtures containing CN and E7 leads to minor stabilisation of BPs, in comparison with ISP mixtures based on COC and E7 (Figure 10). Figure 12 (a) shows the phase transition temperatures of the mixture containing 62.3 wt. % of CN and 31.7 wt. % of nematic E7 as function of concentration of BaTiO$_3$ NPs. The maximum broadening of BPs is observed upon adding of NPs within concentration range from 0.3 to 1.2 wt. %.

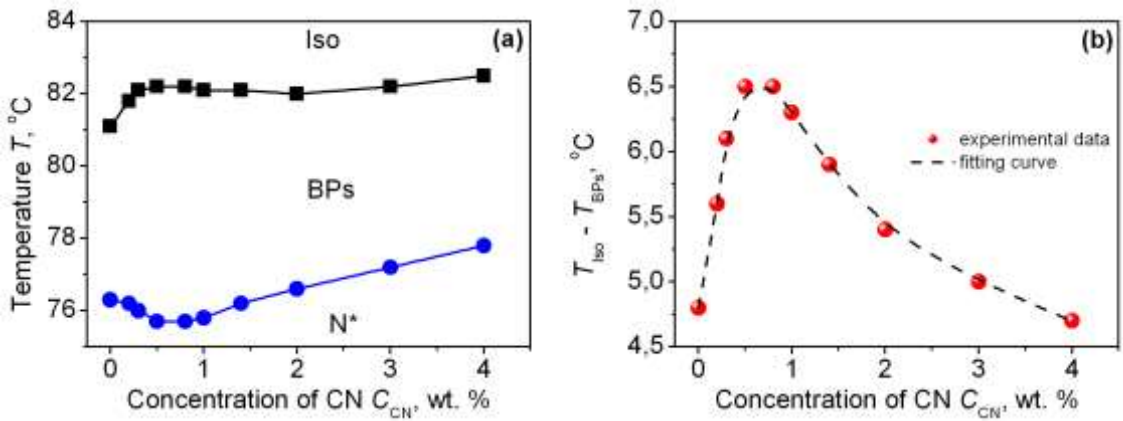

Figure 12. Dependence of the (a) temperature of Iso – BPs (solid black squares) and BPs – N$^*$ (solid blue circles) transitions and (b) temperature difference between Iso and BPs (solid red spheres) during cooling for the E7-CN system consisting of 62.3 wt. % of CN and 31.7 wt. % of nematic E7, on concentration of BaTiO$_3$ NPs.

Figure 12 shows the dependence of temperature Iso - BPs, Iso - N$^*$ and BPs - N$^*$ transitions on concentration of CN of the mixtures doped by 0.5 wt. % of BaTiO$_3$ NPs. As in case of the ISP mixtures containing COC, doping by 0.5 wt. % of BaTiO$_3$ NPs the increase of the deviation from linear dependence $T_{Iso}(C)$ is observed.

In comparison with mixtures having no NPs (Figure 9) the presence of NPs leads to the broadening of temperature range of BPs, which is the reason of sufficient increase in the deviation from linear dependence both on heating and cooling occurs (Figure 13).

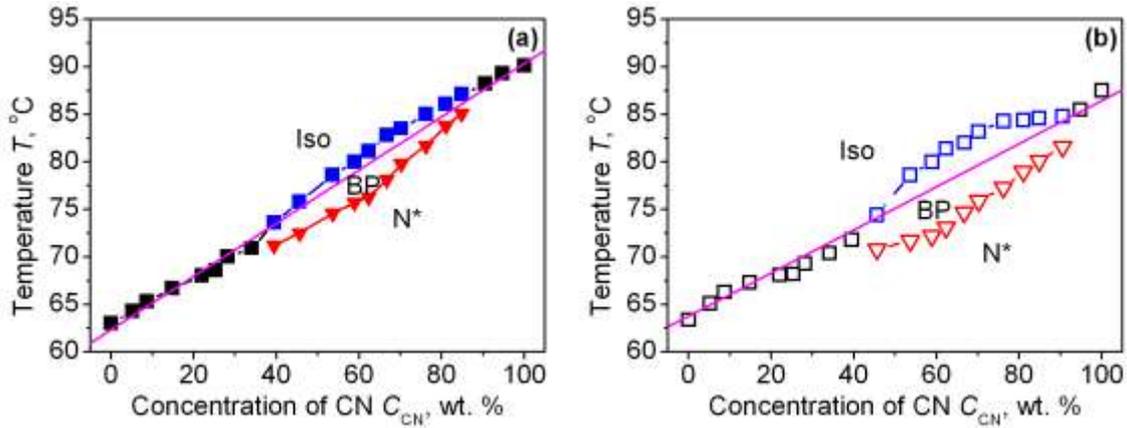

Figure 13. Dependence of the temperature of transitions Iso - BPs (blue squares symbols) and BPs - $N^*$ (red triangles symbols) E7-CN systems on concentration of CN during (a) heating (solid symbols) and (b) cooling (opened symbols). ISP mixtures consisting of certain concentration ratio between CN and E7 and additionally doped by 0.5 wt. % $BaTiO_3$ NPs. The studied E7-CN system contains 62.3 wt. % of CN and 31.7 wt. % of nematic E7. The linear dependence $T(C)$ is shown by solid magenta line.

*3.5 Experimental study of ISP mixtures based on COC and $N_{tb}$-forming mixture*

In this section $N_{tb}$-COC systems containing certain concentration of COC in the $N_{tb}$-forming mixture will be considered to make a comparison between the behavior of conventional 4-cyanobiphenyls and bis(4-cyanobiphenyl-4′-yl) alkanes (CBnCB) with odd n values under similar conditions. We will use the $N_{tb}$-forming mixture recently described in detail elsewhere, [53,54] which consists of two nematic dimers CB7CB, CB6OCB and nematic 5CB in the weight ratio 39:19:42, respectively. Here, the temperature of Iso transition as function of COC concentration ($C_{COC}$) attracted the main attention.

It is also important to note that at certain concentration range of COC, which is approximately from 19 to 90 wt. %, these $N_{tb}$-COC systems are in a specific state characterised by two separated phases (Figure 14). Each of these phases characterised a separated branch of the phase transitions. Speaking of the transition to Iso, then Figure 14 shows the Iso phase transitions for both high temperature (opened black circles) and low temperature (opened green squares) branch of the phase transitions.

Taking into account the temperatures of the phase transitions for the $N_{tb}$ [53] we can conclude that the high temperature branch ($Iso_1$) corresponds to the Iso phase transition of $N_{tb}$-forming mixture. Then the low temperature branch ($Iso_2$) can be attributed to the phase transition of the nematic 5CB. [9]

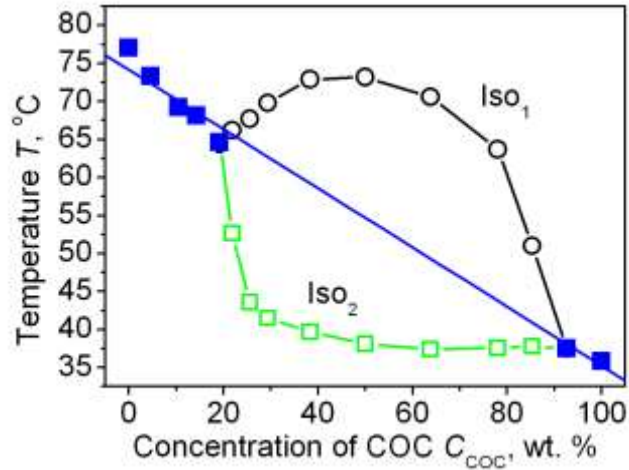

Figure 14. The dependence of temperature of the transition to Iso for systems containing COC and the $N_{tb}$-forming mixture on concentration of COC. Two branches of Iso phase transition at high (opened black circles) and low (opened green squares) temperature are observed within concentration range from 19 to 90 wt. %. The linear dependence $T_{Iso}(C_{COC})$ is shown by solid blue line.

The linear dependence of the $T_{Iso}(C_{COC})$ is shown by the solid blue line (Figure 14). The deviation from linear dependence $T_{Iso}(C_{COC})$ is observed for the $N_{tb}$-COC systems, containing COC within concentration range from 19 to 90 wt. % for which the presence of two branches of phase transitions is observed. For this concentration range the appearance of $S_A$ was also observed (Figure 15(d)).

Figure 15 shows the sequential textures of phase transitions on cooling for the $N_{tb}$-COC system containing COC and $N_{tb}$-forming mixture in weight ratio 1:1. There exist two separated phase transitions for this complex mixture. The sequential independent phase transitions, namely Iso-BPs/$N^*$-$S_A$ are observed.

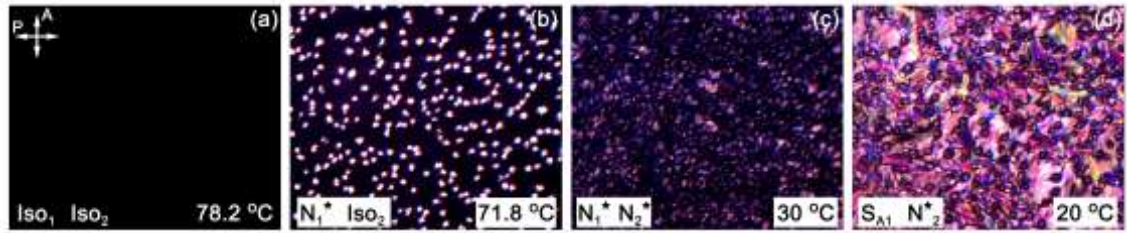

Figure 15. The sequential textures of phase transitions on cooling for the $N_{tb}$-COC system, containing 50 wt. % of COC and 50 wt. % of $N_{tb}$-forming mixture: (a) Iso at 78.2 °C; (b) the appearance of bubble texture ($N^*_1$) of the high temperature branch and $Iso^*_2$ of the low temperature branch at 71.8 °C; (c) $N^*_1$ and $N^*_2$ at 30 °C; (d) the appearance of $S_{A1}$ of the high temperature branch and $N^*_2$ of the low temperature branch at 20 °C. The $N_{tb}$-forming mixture consists of two nematic liquid crystal dimers CB7CB, CB6OCB and nematic 5CB in the weight ratio 39:19:42, respectively.

**Conclusions**

Our idea was to study peculiar features of phase transitions in liquid crystal systems based on nematics and cholesterol esters as a peculiar type of chiral dopants. In particular, we used both conventional nematic mixtures like E7 and an $N_{tb}$-forming mixture comprising CB7CB, CB6OCB and 5CB, as well as smectogenic cholesteryl oleyl carbonate (COC) and cholesteryl nonanoate (CN).

The E7-COC systems in the concentration range of ~ 25 - 80 wt. % COC showed a pronounced increase in the $S_A - N^*$ transition temperature. This effect, which can be considered as a sort of induced smectic phase (ISP) formation, was noticeably stronger than that observed earlier with 5CB, but it nearly vanished with CN, which supported the molecular packing mechanism of ISP formation. As an accompanying effect, it was found that such ISP mixtures showed a certain deviation from linearity of the concentration dependence of temperature Iso and the appearance of the BPs, the concentration range of which clearly correlated with the ISP formation.

The adding of the NPs, in particular, ferroelectric $BaTiO_3$ NPs, to the ISP has led to an increase of the deviation from linear dependence $T_{Iso}(C)$ caused by the stabilisation of BPs (*i.e.* broadening of the temperature range of BPs).

In mixtures based on COC and $N_{tb}$-forming nematic, no induced smectics were observed, but they were characterised by two branches of the Iso transitions at higher and lower temperatures. In this case the deviation from linear dependence of the temperature of Iso transition on concentration of the chiral component is also observed.